\documentclass[final]{aipproc}
\layoutstyle{8x11single}
\usepackage{amssymb}
\begin{document}

\title{Is the Rapid Decay Phase from High Latitude Emission?}


\author{F. Genet}{address={University of Hertfordshire}}

\author{J. Granot}{address={University of Hertfordshire}}

\begin{abstract}
There is good observationnal evidence that the Steep Decay Phase (SDP) that is observed in most \emph{Swift} GRBs is the tail of the prompt emission. The most popular model to explain the SDP is Hight Latitude Emission (HLE). Many models for the prompt emission give rise to HLE, like the popular internal shocks (IS) model, but some models do not, such as sporadic magnetic reconnection events. Knowing if the SDP is consistent with HLE would thus help distinguish between different prompt emission models. In order to test this, we model the prompt emission (and its tail) as the sum of independent pulses (and their tails). A single pulse is modeled as emission arising from an ultra-relativistic thin spherical expanding shell. We obtain analytic expressions for the flux in the IS model with a Band function spectrum. We find that in this framework the observed spectrum is also a Band function, and naturally softens with time. The decay of the SDP is initially dominated by the tail of the last pulse, but other pulses can dominate later. Modeling several overlapping pulses as a single broader pulse would overestimates the SDP flux. One should thus be careful when testing the HLE.
\end{abstract}

\maketitle

\section{Introduction}

Most gamma-ray bursts (GRBs) observed by the \emph{Swift} satellite show an early steep decay phase (SDP) in their X-ray light curve. It is usually a smooth spectral and temporal continuation of the GRB prompt emission, strongly suggesting that it is the tail of the prompt emission \cite{obrienetal06}. It is generally explained by High Latitude Emission (HLE), where at late times the observer still receives photons from increasingly larger angles relative to the line of sight, due to the longer path lenght caused by the curvature of the emitting region. These late photons have a smaller Doppler factor, which results in a steep decay of the flux and in a simple relation between the temporal and spectral indices $\alpha=2+\beta$, where $F_{\nu}(T) \propto T^{-\alpha} \nu^{-\beta}$ \cite{kupa00a}. We test the consistency of HLE with the SDP by modeling the prompt emission as a sum of its individual pulses, including their tails. We calculate the flux for a single emission episode in the framework of internal shocks, and then combine several pulses to model the prompt emission.

\section{Emission of a single pulse}

We consider an ultra-relativistic ($\Gamma \gg 1$) thin (of width $\ll R/\Gamma^2$) spherical expanding shell emitting over a range of radii $R_0 \le R \le R_f \equiv R_0+\Delta R$. The Lorentz factor of the emitting shell is assumed to scales as a power-law with radius, $\Gamma^2 = \Gamma_0^2(R/R_0)^{-m}$, where $\Gamma_0 \equiv \Gamma(R_0)$. In order to calculate the flux received at any time $T$ by the observer we intergrate over the Equal Arrival Time Surface (EATS; \cite{gracota08}), which is the locus of points from which photons that are emitted at a radius $R$, angle $\theta$ from the line of sight and lab frame time $t$ reach the observer at the same observed time $T$. For a shell ejected at an observer time $T_{ej}$, the first photon reaches the observer at a time $T_{ej}+T_0$ with $T_0 = (1+z)R_0/[2(m+1)c\Gamma_0^2]$. We also define $T_f \equiv T_0 (R_f/R_0)^{m+1} = T_0 (1+\Delta R/R_0)^{m+1}$, which is the last time at which photons emitted from the line of sight reach the observer.

We choose for the emission spectrum the phenomenological Band function (Band et al., 1993) spectrum, which generally provides a good fit to the prompt GRB emission. The co-moving peak spectral luminosity is assumed to scale as a power-law with radius, $L'_{\nu'_p} \propto (R/R_0)^a$, where $\nu'_p(R)$ is the peak frequency of the emitted $\nu F_{\nu}$ spectrum. Since Internal shocks is the most popular model for the prompt emission, we consider it for the following. In this framework, several simplifying assumptions can be made: the outflow is expected to be in the coasting phase ($m=0$), and the electrons are expected to be in fast cooling regime. The emission mechanism is assumed to be synchrotron. This leads to $\nu'_p \propto R^{d}$ with $d=-1$, and $L'_{\nu'_p} \propto (R/R_0)^1$, i.e $a = 1$. Then, $T_0 = (1+z)R_0/(2c\Gamma_0^2)$, $T_f = T_0 (1+\Delta R/R_0)$ and the luminosity is
\begin{equation} \label{eq_Lnu_band_DRdiff0}
L'_{\nu'} = L'_0 \left(\frac{R}{R_0}\right)^a S\left(\frac{\nu'}{\nu'_p}\right),   \qquad \qquad
S(x) = e^{1+b_1} \left\{ \begin{array}{ll}
x^{b_1} e^{-(1+b_1)x}  &  x < x_b,\\
x^{b_2} x_b^{b_1-b2} e^{-(b_1-b_2)}  &  x > x_b,
\end{array} \right.
\end{equation}
where S is the normalized Band function, $x \equiv \nu'/\nu'_p$, with $\nu'=(1+z)\nu/\delta$ where $\nu$ is the observed frequency, $x_b = (b_1-b_2)/(1+b_1)$, and $b_1$ and $b_2$ are the high and low energy slopes of the spectrum; $z$ is the redshift of the source and $d_L$ the luminosity distance between the source and the observer. 
We define $\nu_0 = 2\Gamma_0 \nu'_0/(1+z)$, where $\nu'_0 \equiv \nu'_p(R_0)$. One should note that most of the results derived in the following hold only in the model of internal shocks, and not in a more general case. 

\begin{figure}[tb!]
\includegraphics[width=0.33\textwidth, height=0.3\textwidth]{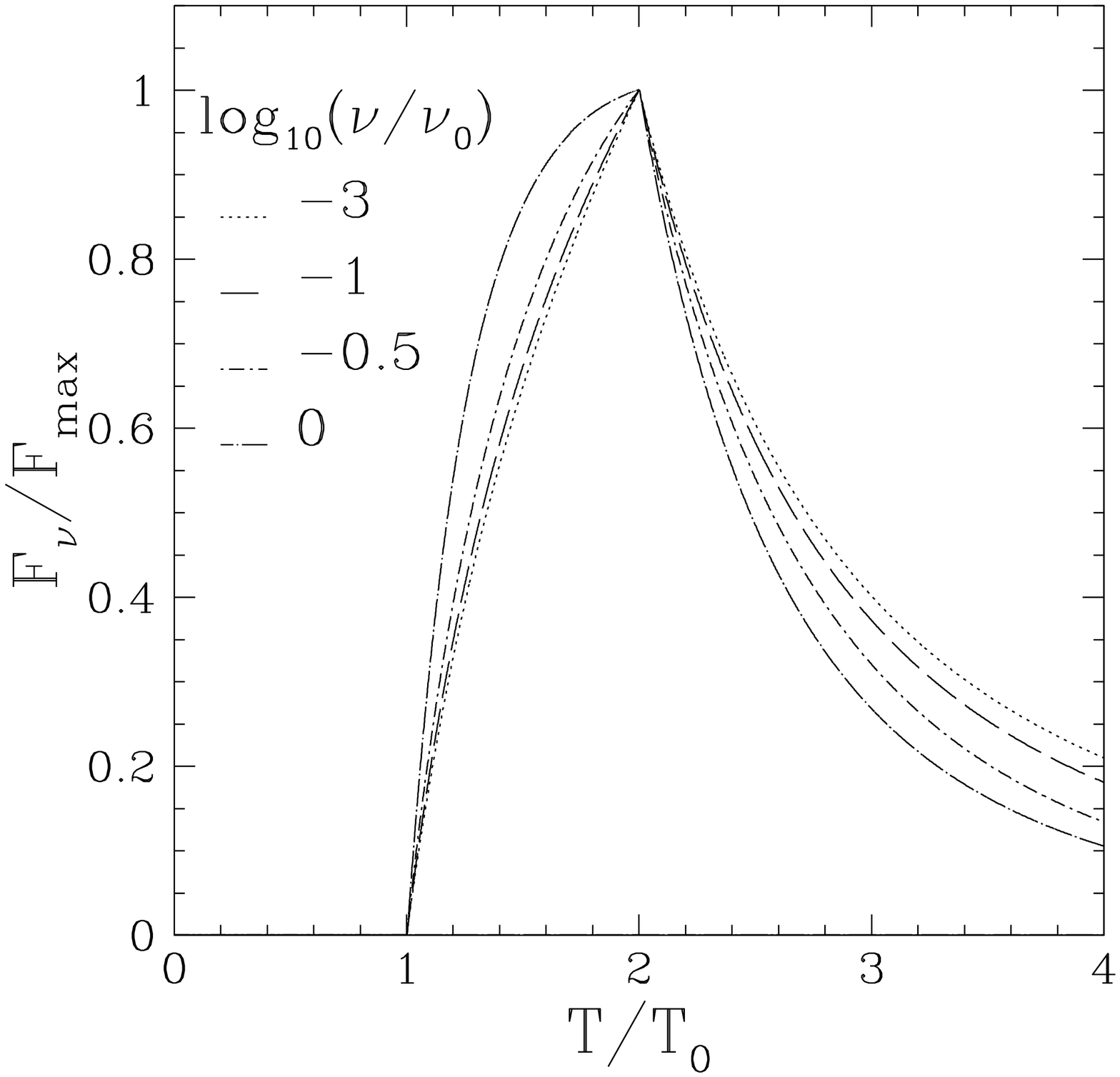}\includegraphics[width=0.33\textwidth, height=0.3\textwidth]{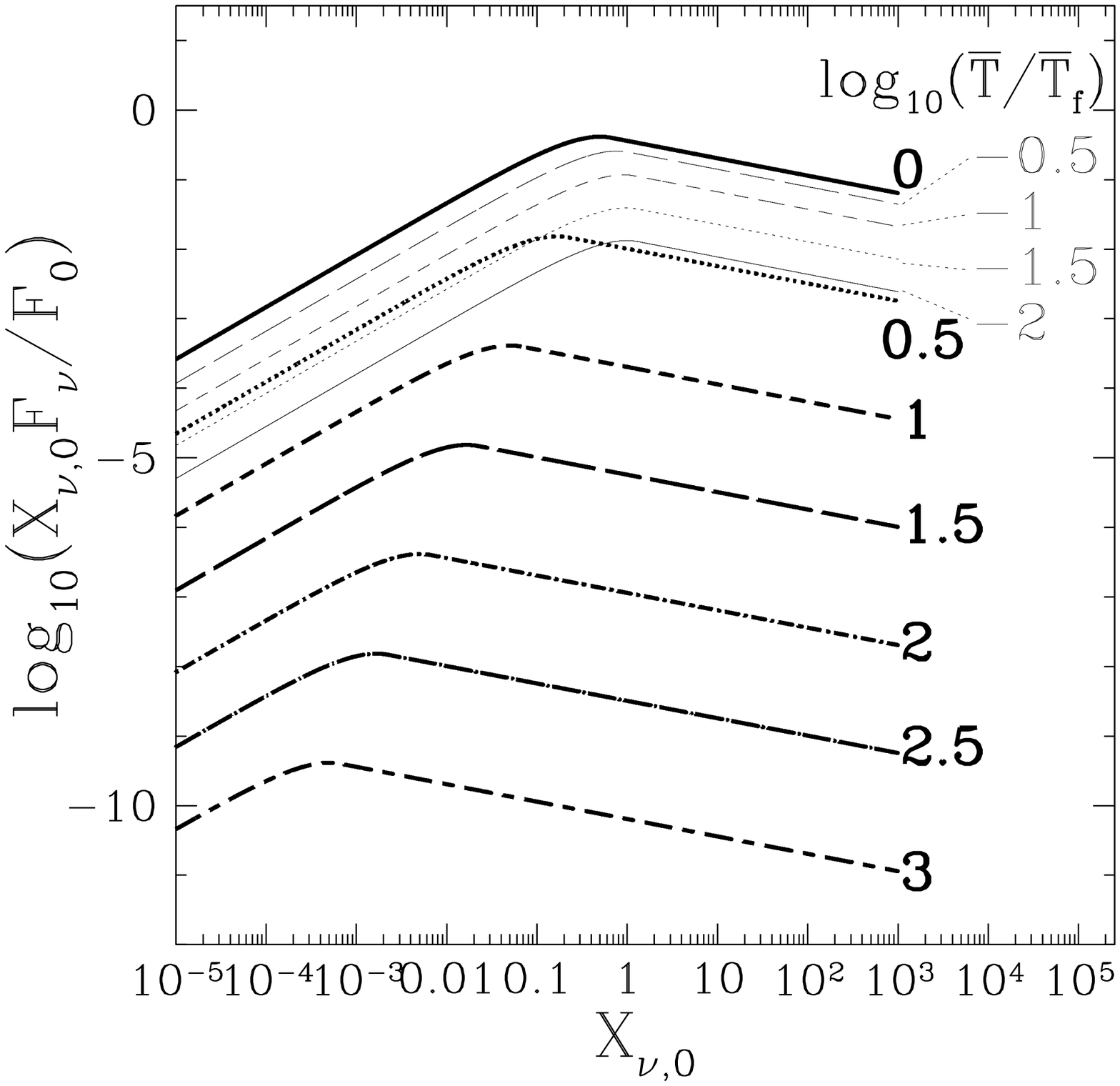}\includegraphics[width=0.33\textwidth, height=0.3\textwidth]{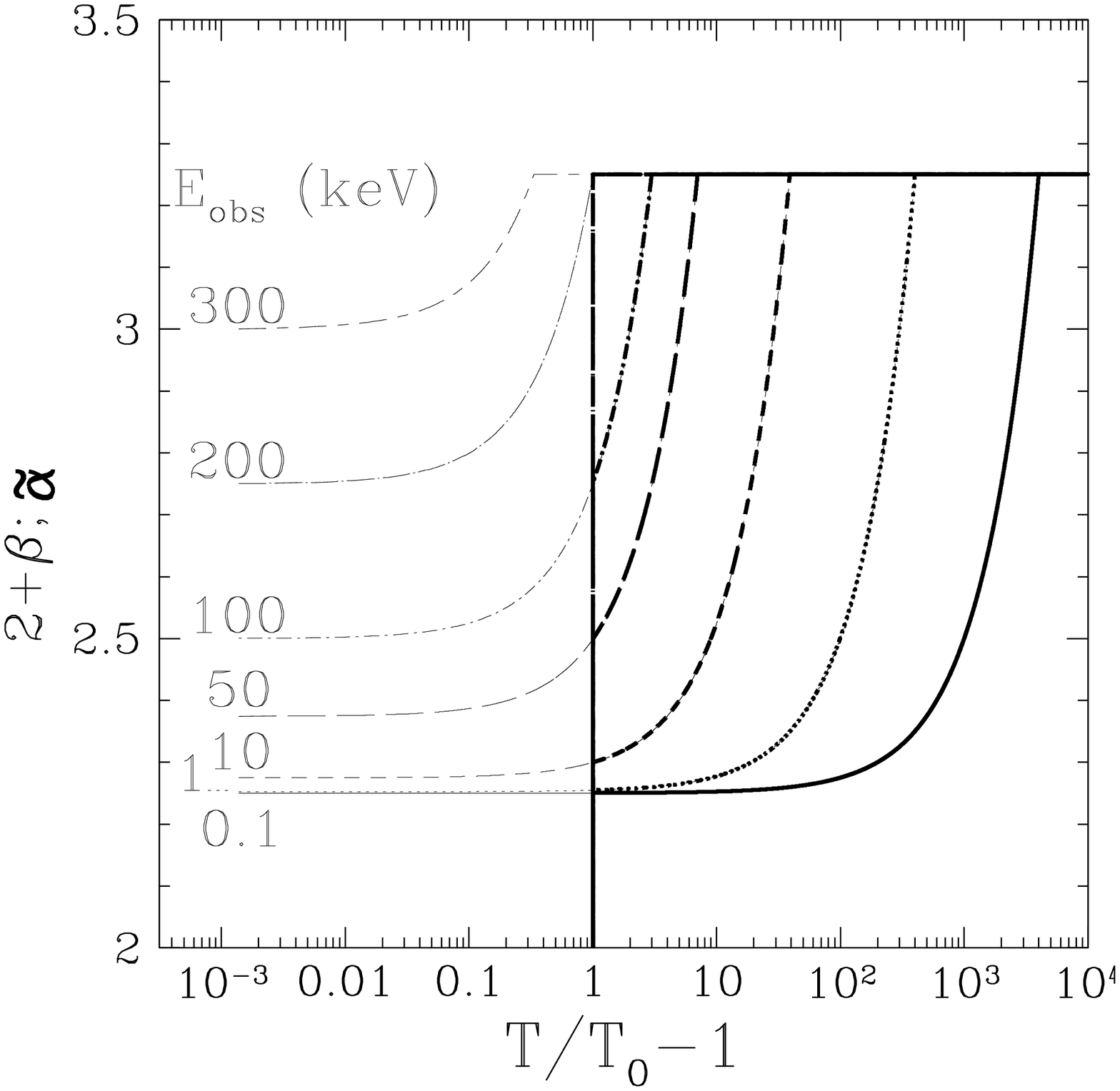}
\caption{{\bf Left:} Evolution of the shape of one normalized pulse with the normalized frequency $\nu/\nu_0$. {\bf Middle:} Evolution of the observed spectrum with time (corresponding to the values of $\bar{T}/\bar{T_f}$ written near each spectrum). The thin lines correspond to the rising part of the pulse, the thick lines to the decaying part of the pulse. $\Delta R/R_0=1$. $\nu/\nu_0(T_0) = 1$. {\bf Right}: Comparison of the evolution of the spectral ($2+\beta$; thin lines) and temporal ($\alpha$; thick lines) slopes at fixed observed frequencies (for ${E'}_0=0.5$ keV and $\Gamma_0=300$, so that $E_{0,obs}=300$ keV).} \label{fig_spectrum}
\end{figure}

The observed flux is then (in the framework of internal shocks):
\begin{eqnarray}\label{eq_bandDR_finalcondensed}
F_{\nu}(T \geq T_{\rm ej}+T_0) = F_0
\left(\frac{T-T_{\rm ej}}{T_0}\right)^{-2} 
\left[\left(\frac{\min(T-T_{\rm ej},T_f)}{T_0}\right)^3-1\right] 
S\left(\frac{\nu}{\nu_0}\frac{T-T_{\rm ej}}{T_0}\right)\ ,
\end{eqnarray}
where $F_0 \equiv (1+z)L_0/(12\pi d_L^2)$. Figure \ref{fig_spectrum} (left panel) shows the variation of a pulse shape with the normalized frequency $\nu/\nu_0$. Different shapes can be obtained, form spiky to rouder. For $m = 0$ and $d = -1$ the observed spectrum is a pure Band function, just like the emitted spectrum (see middle panel of Fig. \ref{fig_spectrum}), where the observed peak frequency of the $\nu F_\nu$ spectrum decreases with time as $\nu_p = \nu_0/\tilde{T}$, and $\tilde{T} = (T-T_{\rm ej})/T_0 = 1 + \bar{T}$. This corresponds to a softening of the spectrum with time (Fig. \ref{fig_spectrum} right panel) which agrees with observations. On the same panel we compare the evolution of the instantaneous spectral slope $\beta \equiv -d \log F_{\nu}/d \log \nu$ with the temporal slope $\tilde{\alpha} \equiv -d\log F_\nu/d\log\tilde{T}$, where $\tilde{T} = (T-T_{\rm ej})/T_0 = 1+\bar{T}$: we can see that the HLE relation $\tilde{\alpha} = 2+\beta$ is valid as soon as $\bar{T}>\bar{T}_f$. One should be careful that this is true only in the framework of internal shocks model, and with this definition of the temporal slope (for exemple, $\bar{\alpha} \equiv -d \log F_{\nu}/d \log \bar{T}$, which is another definition of the temporal slope, approaches $2+\beta$ only at late time).


\section{Combining pulses to obtain the prompt emission}

\begin{figure}[tb!]
\includegraphics[width=0.45\textwidth, height=0.4\textwidth]{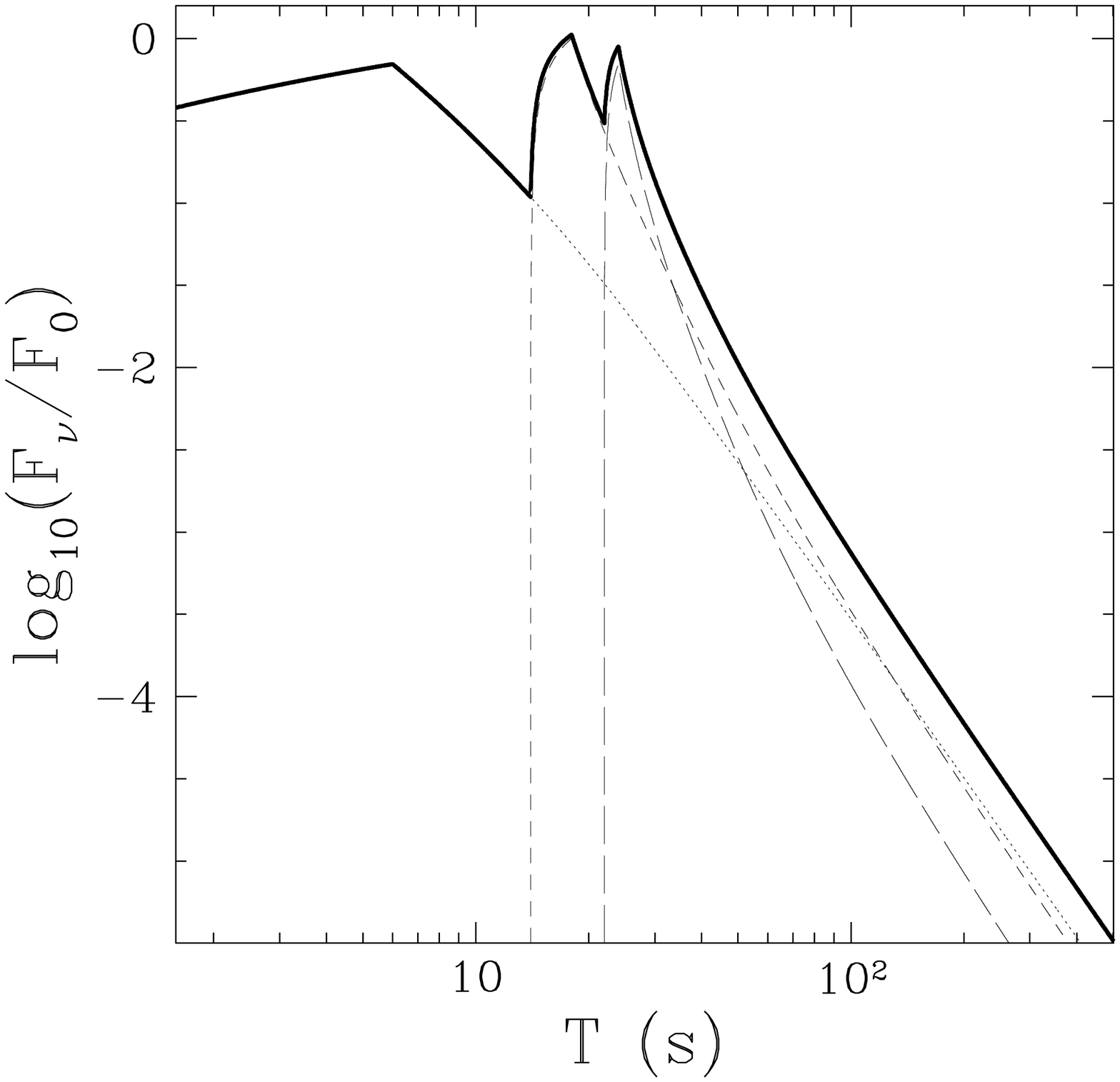}\includegraphics[width=0.45\textwidth, height=0.4\textwidth]{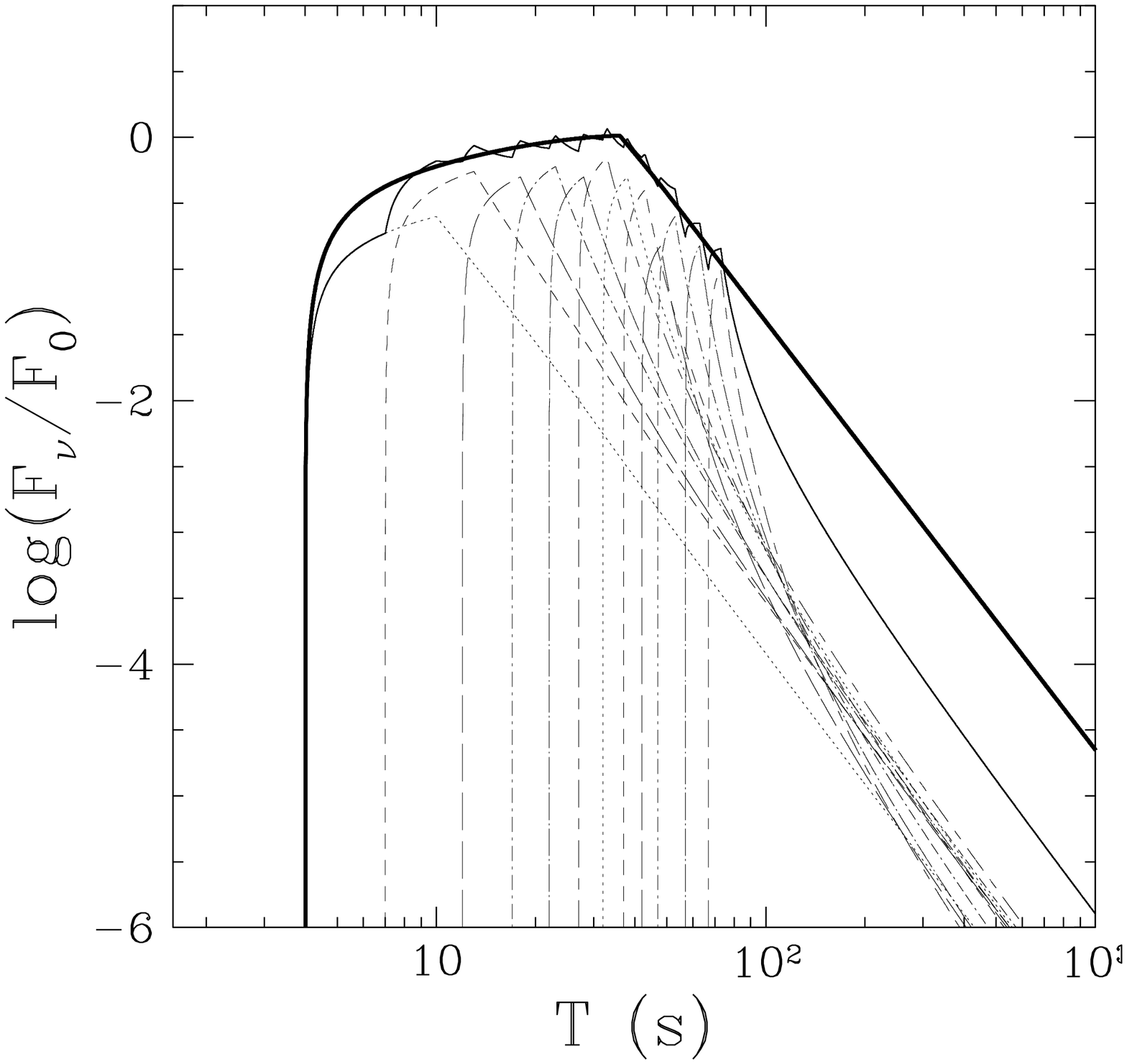}
\caption{
{\bf Left:} \emph{Exemple of a prompt emission consisting of three pulses with $T_{\rm ej} = -1\;$s, $13\;$s, $21\;$s, $T_0 = 2\;$s for all three pulses, $\Delta R/R_0 = 3$, $2$, $1$, and $F_{\rm peak}/F_0 = 0.7$, $1$, $0.7$. Thin non-solid lines represent individual pulses, while the thick solid line shows the total prompt emission.} {\bf Right:} \emph{Comparison between a fit with several pulses ($12$ here) and a fit with one broad pulse. Thin non solid lines shows each individual pulse, the thin solid line shows the total prompt emission, and the thick solid line shows a possible fit with one broad pulse. The normalized frequency is $\nu/\nu_0 = 0.1$. Both panels are in logarithmic scale.}} \label{fig_compfit}
\end{figure}

Within our model, the prompt emission is the sum of independent pulses, and the SDP is thus the sum of the tails of these pulses. For a prompt emission composed of several equal pulses, at late time the contribution of each pulse is equal, and the temporal slope just after the peak of a pulse increases with its ejection time $T_{ej}$. When varying several parameters among the different pulses, the late time flux ratio of the pulse tails is the ratio of their $F_{peak}T_f^{2+\beta}$. Just after the peak of the last pulse, the SDP is dominated by the last pulse. This shows that several pulses can dominate the SDP at different times, as one can see in the left panel of Fig. \ref{fig_compfit}. Therefore, one should be careful to consider this when studying the temporal and spectral behavior of the SDP.

Figure \ref{fig_compfit} (right panel) shows what can happen if, because of noisy data or coarse time bins, a prompt emission (thin solid line) which is actually composed by several pulses is fitted by one broad pulse (thick solid line): the fit would give a tail with the same temporal slope at late time than the actual prompt tail, but with no higher temporal slopes just at the end of the prompt, the whole tail of this broad pulse being close to a power law. Moreover, this overestimates the flux of the SDP. It is important to keep this in mind when confronting such a model with actual data.

\section{Conclusion}

We have outlined a model for the prompt emission and its tail. This model contains a restricted number of free parameters, $10$ per pulse: $a$, $m$, $d$, $F_0$, $b_1$, $b_2$, $E_0(T_0)$, $T_0$, $T_f$ and $T_{ej}$. In the case for internal shocks, this can be reduced to $7$: $m=0$; $d=-1$ and $a=1$; as in this framework $\Delta R \sim R_0$ is expected, one can fix $F_f/T_0 = 1+ \Delta R/R_0 = 2$. For a prompt emission with $N$ pulses, the total number of free parameters can be further reduced to $3(N+1)$, instead of $6N$, as we expect the Band function parameters ($b_1$, $b_2$ and $E_0(T_0)$) to be similar for all pulses.

The shape of a pulse can vary considerebly in our model, from very spiky to rounder, which qualitatively reproduces the observed diversity. The observed spectrum is a pure Band function as the emitted one in the case of internal shocks, and our model naturally produces a softening of the spectrum, as is observed.

When combining several pulses to model the prompt emission, the Steep Decay Phase is initailly dominated by the last pulse, and is dominated at late times by the pulse with the largest $F_{peak}T_f^{2+\beta}$ (essentially the widest pulse, except if there is a large difference of flux between the pulses), but can be dominated by other pulses in between.

When fitting data, one should be careful not to consider several overlapping pulses as a single broad pulse, which would lead to an overestimate of the prompt tail flux and a misinterpretation of the steep decay phase.



\end{document}